\documentclass{article}
\usepackage{spconf,graphicx,hyperref}
\usepackage{amsmath,amssymb,bm}
\usepackage{multirow,booktabs,enumitem}
\hypersetup{hidelinks}

\title{CochlearMark: Audio Watermarking via \\Phase Locked Auditory Nerve Representations}

\twoauthors
{Qin Liu\sthanks{qin.liu@epfl.ch}, Strahinja Nikolic\sthanks{The first and second authors performed this work while at Logitech}}
	{EPFL\\
	1015 Lausanne, Switzerland}
{Milos Cernak}
	{Logitech\\
	1015 Lausanne, Switzerland}

\begin{document}
%
\maketitle
\begin{abstract}

Modern audio watermarks achieve strong robustness with learned encoder-decoder models, but their behavior under new processing is established empirically. CochlearMark is introduced as a training-free audio watermark that uses auditory-nerve phase-locking as its feature representation. A small waveform perturbation is optimized against a keyed random projection of this representation, and the same projection is used for detection, with no trained encoder or decoder. Because phase-locking represents temporal fine structure, the resulting robustness has a direct physiological interpretation: the mark is retained when this structure is preserved and lost when it is disrupted or rebuilt. Across speech, music, and environmental audio, CochlearMark achieves signal-edit robustness and imperceptibility comparable to trained neural baselines while requiring only a lightweight computational footprint. Among the evaluated training-free methods, CochlearMark is the only one to retain substantial detectability under neural-codec re-encoding.

\end{abstract}

\section{Introduction}
\label{sec:intro}
Audio watermarking embeds an imperceptible, detectable signature in a signal to trace provenance and authenticate audio. Recent methods use neural networks with learned models for watermark embedding and detection~\cite{roman2024audioseal,chen2023wavmark,pavlovic2025aware,singh2024silentcipher,li2025voicemark,zhou2024wmcodec}. AudioSeal jointly trains a watermark generator and detector for localized detection~\cite{roman2024audioseal}, while WavMark learns multi-bit watermark embedding and recovery~\cite{chen2023wavmark}. Other systems extend this general learned watermarking framework to different robustness and generation settings~\cite{li2025voicemark,zhou2024wmcodec}.

Robustness to signal processing has also become an explicit design objective for audio watermarking. SilentCipher incorporates differentiable distortions and pseudo-differentiable compression during training to improve robustness to audio codecs~\cite{singh2024silentcipher}. AWARE replaces a fixed stack of simulated distortions with adversarial optimization, targeting robustness to unseen edits~\cite{pavlovic2025aware}. These methods achieve strong robustness and high imperceptibility, but robustness still emerges from learned models and training-time objectives. Behavior under new processing therefore remains empirical and must be established by evaluation~\cite{kumara2026fragile,yao2025overwriting}.

Training-free methods take a different approach. Classical designs such as spread spectrum, quantization index modulation, and patchwork use fixed embedding and detection rules in the waveform or a transform domain~\cite{cox1997spread,chen2001qim,bender1996techniques}. More recently, Hidden in Plain Tokens proposed a gradient-free watermark for synthetic audio by exploiting redundancy in a learned discrete token representation~\cite{milis2026hidden}. This achieves robustness without training, but operates within the discrete representation of a generative audio system.

A gap therefore remains for training-free watermarking applied directly to waveform audio. Such a method needs a feature representation that is robust to signal edits without learning this behavior from data. The auditory nerve provides a candidate representation. Auditory-nerve fibers fire at a consistent phase of the stimulus fine structure, a property called phase-locking~\cite{goldberg1969response,joris2004neural,reichenbach2012frequency}. Phase-locking represents the temporal fine structure of the carrier. Many conventional signal edits largely preserve this structure, whereas processing that disrupts or rebuilds the waveform alters it.

CochlearMark builds on this idea using a differentiable model of the auditory nerve. The main contributions are:
\begin{itemize}[leftmargin=2em, labelsep=0.5em]

\item Auditory-nerve phase locking is introduced as the watermark feature space, giving a physiologically interpretable robustness boundary determined by preservation of temporal fine structure.

\item A keyed random projection serves as detector and embedding target, without training an encoder or detector.

\item CochlearMark achieves signal-edit robustness and imperceptibility comparable to trained neural baselines with a lightweight computational footprint. Among the evaluated training-free methods, CochlearMark is the only one to retain substantial detectability under neural-codec re-encoding.

\end{itemize}

\section{Method}
\label{sec:method}

\subsection{Framework overview}
\label{ssec:overview}

\begin{figure*}[ht]
  \centering
  \includegraphics[width=0.9\textwidth]{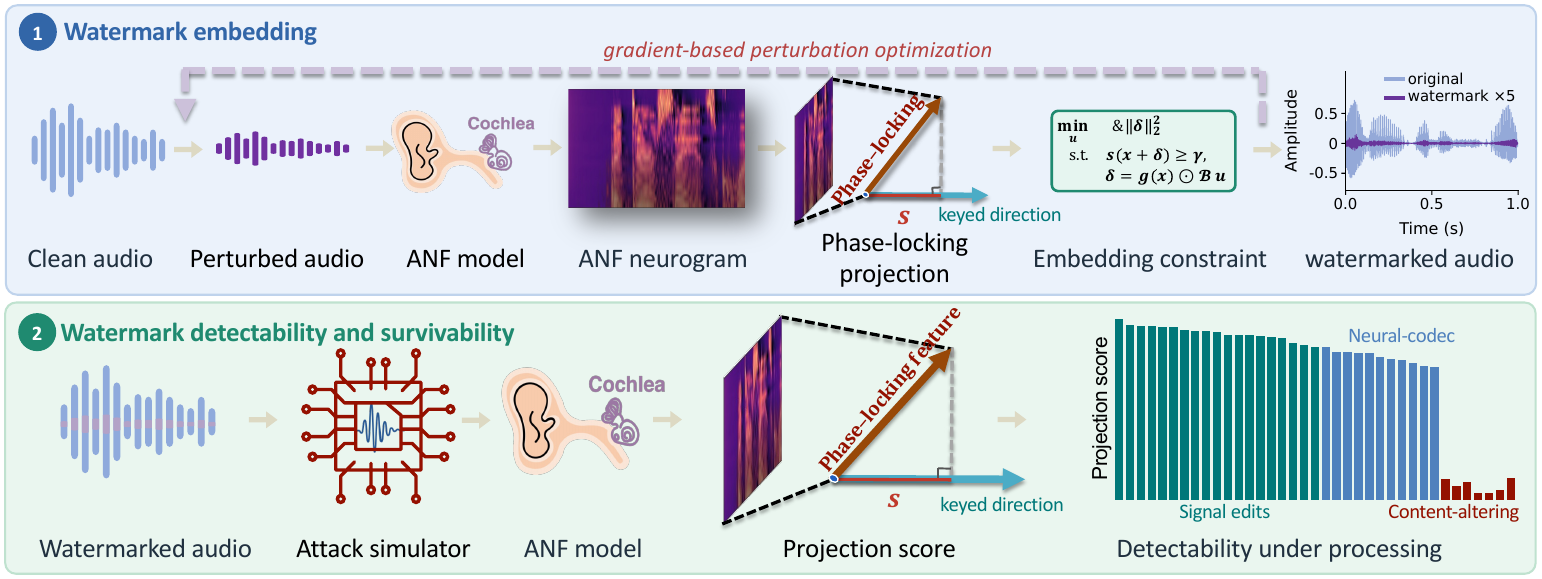}
  \vspace{-10pt}
  \caption{Overview of CochlearMark. (1) Watermark embedding: a waveform perturbation is optimized through the differentiable ANF model so that the phase-locking feature yields a keyed projection score above the embedding threshold. Once the constraint is satisfied, the optimized signal is retained as the watermarked audio. The waveform example at right shows the original audio and the embedded perturbation, with the perturbation amplified by $5\times$ for visibility. (2) Watermark detectability and survivability: the watermarked audio is processed by an attack simulator and re-evaluated using the same ANF model and keyed phase-locking projection. The schematic profile illustrates high detectability under signal edits, decreasing detectability under neural-codec re-encoding, and low detectability under content-altering transformations.}

  \label{fig:framework}
\end{figure*}

CochlearMark embeds the watermark as a small additive perturbation in the waveform. As illustrated in Fig.~\ref{fig:framework}, the audio signal is passed through a differentiable auditory-nerve-fiber (ANF) model. The model produces ANF firing rates over a set of characteristic frequency channels and time bins. From these firing rates, a phase-locking feature is projected onto a keyed direction to give a watermark detection score. Embedding optimizes the perturbation to raise this score above a detection threshold, whereas detection applies the same projection to the test signal and compares the score with the threshold.

\subsection{Cochlear front-end and keyed detector}
\label{ssec:detector}

The cochlear front-end is a differentiable ANF model\footnote{Differentiable ANF model by M. Saddler, \url{https://github.com/msaddler/auditory_machine_learning}.} based on the auditory-periphery model of Bruce \textit{et al.}~\cite{Bruce2018}. It includes a gammatone filterbank, half-wave rectification, and an inner-hair-cell (IHC) low-pass stage. The IHC low-pass stage limits the temporal fluctuations represented in the firing rate, concentrating phase-locking information at frequencies below the cutoff frequency $f_{\mathrm{IHC}}$.

The front-end, denoted by $\Phi$, maps an $N$-sample waveform $x \in \mathbb{R}^{N}$ to an auditory-nerve neurogram $z = \Phi(x) \in \mathbb{R}^{C \times T}$ with $C$ frequency channels and $T$ time frames. The entry $z_{c,t}$ denotes the firing rate of channel $c$ with characteristic frequency $f_c$ at frame $t$. The stochastic synapse stage of Bruce \textit{et al.} is replaced by a sigmoid rate-level function to make the front-end differentiable with respect to the waveform. This allows gradients to propagate from the detection score back to the waveform during watermark embedding.

The degree of phase-locking in each channel is quantified by the vector strength of the neurogram at the channel characteristic frequency:
\begin{equation}
  \psi_c = \frac{\left| \sum_{t=1}^{T} z_{c,t}\, e^{-\mathrm{i} 2\pi f_c t / f_s} \right|}{\sum_{t=1}^{T} z_{c,t}} \in [0,1] .
  \label{eq:plv}
\end{equation}
where $f_s$ is the neurogram frame rate. A vector strength close to $1$ indicates that frequency channel $c$ fires at a consistent phase of the $f_c$ cycle, and a value near $0$ indicates weak phase-locking. Stacking the vector strengths of all channels gives the phase-locking feature $\bm{\psi} = [\psi_1, \dots, \psi_C]^\top \in [0,1]^C$.

The phase-locking feature is projected onto a keyed direction $\bm{d}$ to obtain the watermark detection score:
\begin{equation}
  s = \langle \bm{\psi},\, \bm{d} \rangle ,
  \label{eq:score}
\end{equation}
where $\langle\cdot,\cdot\rangle$ denotes the inner product, and the keyed direction $\bm{d} \in \mathbb{R}^C$ is generated from the detection key and normalized to unit length. A signal is declared watermarked when $s > \gamma$, with the threshold
\begin{equation}
  \gamma = \mu_{\mathrm{clean}} + \kappa\, \sigma_{\mathrm{clean}},
  \label{eq:gamma}
\end{equation}
where $\mu_{\mathrm{clean}}$ and $\sigma_{\mathrm{clean}}$ are the mean and standard deviation of the scores over a set of clean clips, and $\kappa$ is a fixed threshold multiplier. The detector has no trainable parameters; the same projection serves the watermark embedding optimization and detection.

\subsection{Watermark embedding}
\label{ssec:embed}

The watermark is embedded as an additive perturbation $\delta \in \mathbb{R}^{N}$, giving the marked audio $y = x + \delta$. For each clip, embedding is formulated as
\begin{equation}
\begin{aligned}
  \min_{u}\quad & \lVert \delta \rVert_2^2 \\
  \text{s.t.}\quad & s(x+\delta) \ge \gamma, \\
  & \delta = g(x) \odot \mathcal{B}\,u ,
\end{aligned}
\label{eq:embed}
\end{equation}
where $u \in \mathbb{R}^{N}$ is the optimization variable, a waveform of the same length as $x$ that the solver is free to set, and $\lVert\cdot\rVert_2$ is the Euclidean norm, so $\lVert\delta\rVert_2^2$ is the perturbation energy. The operator $\mathcal{B}$ restricts $u$ to a frequency band $[f_\ell, f_b]$, removing energy below $f_\ell$ and above $f_b$. The envelope $g(x) \in [0,1]^N$ is a smoothed amplitude envelope of the clean signal that suppresses the perturbation in silent regions. The operator $\odot$ denotes element-wise multiplication. This parameterization keeps the perturbation band-limited and gated throughout optimization. The detection constraint requires the marked signal to reach the calibrated threshold $\gamma$ in Eq.~\ref{eq:gamma}. The band-limit restricts the perturbation to the frequency range where phase-locking information is concentrated, while the amplitude gate concentrates the perturbation in active signal regions and suppresses it in silence.

The constrained problem in Eq.~\ref{eq:embed} is approximated along the initial score-gradient direction. The detection margin along this one-dimensional path is modeled by a quadratic function, determined from one forward-backward pass and one additional forward probe. Its smallest nonnegative root gives a closed-form estimate of the minimum-energy feasible step along the selected direction. If the threshold is not reached, up to three Adam steps~\cite{kingma2015adam} are applied for correction. The auditory front-end and keyed detector remain fixed throughout embedding.

\section{Experimental evaluation}
\label{ssec:setup}

\subsection{Experimental setup}

CochlearMark is evaluated on three corpora spanning speech, music, and environmental audio: LibriSpeech~\cite{panayotov2015librispeech} for speech (test-clean and test-other), GuitarSet~\cite{xi2018guitarset} for solo-instrument music, and Clotho~\cite{drossos2020clotho} for environmental audio. The evaluation comprises 6{,}964 clips in total. The CochlearMark detection threshold is calibrated on 200 clips from the LibriSpeech dev-clean subset, disjoint from all evaluation data. Four baselines are included: two trained neural watermarks, AudioSeal~\cite{roman2024audioseal} and WavMark~\cite{chen2023wavmark}, and two training-free classical methods, Patchwork (audiowmark)~\cite{audiowmark} and spread spectrum~\cite{cox1997spread}. Cross-method comparisons use the same source clips on each corpus.

Watermark robustness is evaluated by processing each marked signal before detection. The operations are grouped into four categories. Signal edits include resampling to $8$ and $4$~kHz, additive noise, low-pass and high-pass filtering, requantization, MP3 and Opus coding, equalization, all-pass phase scrambling, reverberation, echo, $\mu$-law companding, packet-loss dropout, $5$-ms time shifting and $10\%$ cropping, and a VoIP-style noise-suppression chain based on DeepFilterNet~\cite{schroter2022deepfilternet}. Neural-codec re-encoding is evaluated with EnCodec~\cite{defossez2023encodec} across multiple bitrates. Content-altering transformations include pitch shifting, voice conversion with FreeVC~\cite{li2023freevc}, and text-to-speech resynthesis with XTTS~\cite{casanova2024xtts}. Finally, a removal category evaluates an adversarial attempt to strip the watermark rather than signal transformation, using the recent watermarks-remover tool\footnote{\url{https://github.com/guillaumemeyer/watermarks-remover}}; its remix chain applies $1.08\times$ time stretching, a $+2$-semitone pitch shift, a $35$~Hz--$14$~kHz band limit with $8$~dB treble attenuation, and AAC re-encoding at $96$~kbps.

Detection on unattacked audio is summarized by ROC-AUC and the true-positive rate at $1\%$ false-positive rate (TPR@1\% FPR). Imperceptibility is measured between the marked and clean signals using SNR, PESQ~\cite{rix2001pesq}, and Zimtohrli~\cite{alakuijala2025zimtohrli}. Robustness is summarized by the per-attack survival rate (TPR@1\% FPR), defined as the fraction of marked clips that remain detected after processing.

The front-end uses $C = 36$ gammatone channels on the equivalent-rectangular-bandwidth (ERB) scale from $80$ Hz to $6$ kHz, an IHC low-pass cutoff $f_{\mathrm{IHC}} = 3$ kHz, an internal rate of $20$ kHz, and a neurogram frame rate $f_s = 10$ kHz. The detection feature is the $36$-dimensional phase-locking value, and the keyed direction is generated from a fixed random seed. The threshold multiplier is set to $\kappa = 3.5$. The perturbation is band-limited to $[f_\ell, f_b] = [300\,\mathrm{Hz}, 2.5\,\mathrm{kHz}]$. Embedding uses the quadratic solver described in Sec.~\ref{ssec:embed}. The neurogram frame rate is fixed at $f_s=10$ kHz ($0.1$-ms time bin), independent of clip duration. Recordings are standardized to $4.0$ s, balancing longer-duration evaluation against computational cost. The perturbation is added back to the signal at its native sample rate, preserving content above the watermark band. No component is trained; only the per-clip perturbation variable is optimized.

\begin{figure*}[ht]
  \centering
  \includegraphics[width=\textwidth]{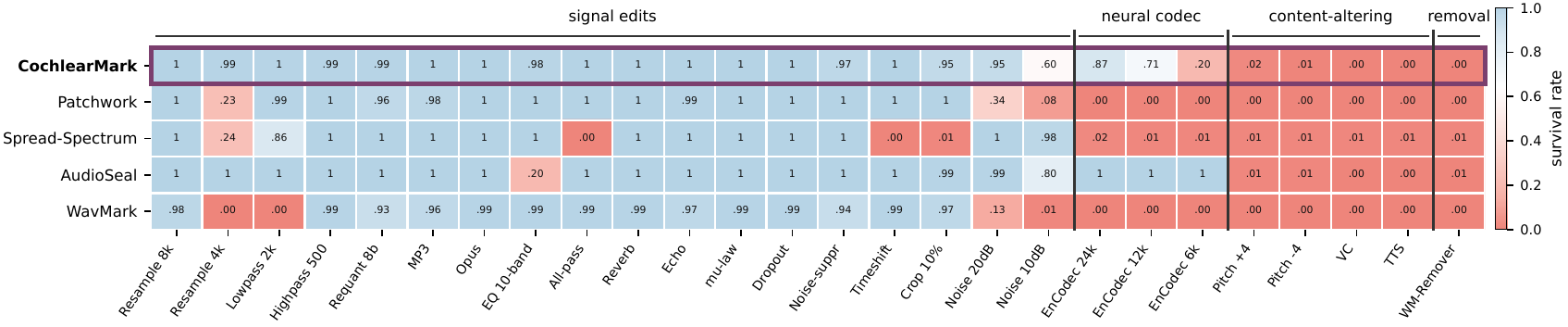}
  \vspace{-25pt}
  \caption{Per-attack watermark survival (TPR@1\% FPR) pooled across LibriSpeech, GuitarSet, and Clotho (6{,}964 clips per method), spanning signal edits, neural-codec re-encoding, content-altering transformations, and a watermark-removal tool.}
  \label{fig:survival}
\end{figure*}

\subsection{Experimental results}

\begin{table}[t]
\centering
\caption{Imperceptibility and detection on unattacked audio.}
\label{tab:quality}
\setlength{\tabcolsep}{3.9pt}
\renewcommand{\arraystretch}{1.1}
\footnotesize
\begin{tabular}{@{}ll ccc cc@{}}
\toprule
Corpus & Method & SNR & PESQ & Zimtohrli & AUC & TPR@1\% \\
\midrule
\multirow{5}{*}{\shortstack[l]{Libri-\\Speech}}
 & \textbf{CochlearMark} & 31.0 & 3.91 & 4.50 & 1.000 & 1.000 \\
 & AudioSeal & 27.5 & 4.41 & 4.94 & 1.000 & 1.000 \\
 & WavMark   & 36.7 & 4.14 & 4.39 & 1.000 & 1.000 \\
 & Patchwork        & 24.6 & 4.52 & 4.91 & 1.000 & 1.000 \\
 & Spread-Spectrum  & 33.4 & 3.56 & 3.91 & 1.000 & 1.000 \\
\midrule
\multirow{5}{*}{\shortstack[l]{Guitar-\\Set}}
 & \textbf{CochlearMark} & 31.6 & 3.79 & 4.76 & 0.987 & 0.975 \\
 & AudioSeal & 24.7 & 3.87 & 4.90 & 1.000 & 1.000 \\
 & WavMark   & 35.8 & 3.88 & 4.33 & 1.000 & 1.000 \\
 & Patchwork        & 26.1 & 4.29 & 4.87 & 1.000 & 1.000 \\
 & Spread-Spectrum  & 36.8 & 3.51 & 4.13 & 1.000 & 1.000 \\
\midrule
\multirow{5}{*}{Clotho}
 & \textbf{CochlearMark} & 31.3 & 4.30 & 4.82 & 0.995 & 0.990 \\
 & AudioSeal & 23.5 & 4.34 & 4.92 & 1.000 & 1.000 \\
 & WavMark   & 37.0 & 4.39 & 4.78 & 0.979 & 0.958 \\
 & Patchwork        & 25.1 & 4.53 & 4.91 & 1.000 & 1.000 \\
 & Spread-Spectrum  & 33.7 & 4.37 & 4.79 & 1.000 & 1.000 \\
\bottomrule
\end{tabular}
\end{table}

Table~\ref{tab:quality} reports imperceptibility and detection on unattacked audio. Across the three corpora, CochlearMark remains within the imperceptibility ranges of the evaluated baselines, while detection stays high, with ROC-AUCs of $0.987$--$1.000$ and TPR@1\% FPRs of $0.975$--$1.000$. Table~\ref{tab:compute} shows that this performance is achieved with a compact computational footprint: CochlearMark uses a 145~KB model and 120~MB peak memory, below the neural model baselines, while both embedding and detection are faster than real time on CPU.

Fig.~\ref{fig:survival} reports the per-attack survival rate (TPR@1\% FPR). Under most signal edits, CochlearMark remains detectable at rates comparable to the evaluated baselines. Survival decreases as the level of additive noise increases. Neural-codec re-encoding produces a bitrate-dependent decline, with EnCodec survival decreasing from $0.87$ at $24$ kbps to $0.20$ at $6$ kbps. Survival is also near zero after the content-altering transformations and after the removal chain, which removes the mark from all five methods.

\begin{table}[t]
\centering
\caption{Computational cost per second of audio on a single CPU (Intel Core i7-6850K, 6\,cores @ 3.6\,GHz). Size is the on-disk footprint and Memory is the peak RAM during processing; in the Compute and Latency columns each entry is \mbox{embed\,/\,detect}. GFLOPs are device-independent.}
\label{tab:compute}
\setlength{\tabcolsep}{3.7pt}
\renewcommand{\arraystretch}{1.15}
\footnotesize
\begin{tabular}{@{}l c c c c@{}}
\toprule
Method & Size & Memory & Compute (GFLOPs) & Latency (ms) \\
\midrule
\textbf{CochlearMark} & 145\,KB & 120\,MB & $\sim$1.0 / 0.13 & 39 / 4.4 \\
AudioSeal        & 93\,MB   & 906\,MB & 2.70 / 1.50               & 38 / 21    \\
WavMark          & 9.9\,MB  & 879\,MB & 80.6 / 913                & 282 / 4607 \\
Patchwork        & 16\,MB   & 260\,MB & $\sim$0.005 / $\sim$0.005 & 7.0 / 78   \\
Spread-Spectrum  & $\sim$10\,KB & 147\,MB & 0.0015 / 0.0007        & 0.5 / 0.3  \\
\bottomrule
\end{tabular}
\end{table}

\section{Discussion}
\label{sec:discussion}

The survival pattern in Fig.~\ref{fig:survival} follows from the phase-locking representation. Signal edits such as coding, filtering, and reverberation largely preserve the temporal fine structure, whereas low-bitrate neural-codec re-encoding, voice conversion, and text-to-speech rebuild the waveform and disrupt the original phase pattern. Pitch shifting changes the frequency-to-channel correspondence and similarly reduces the keyed projection. The resulting robustness boundary is therefore determined by the auditory representation.

EnCodec illustrates that content preservation is not sufficient for watermark preservation: at low bitrates the decoder reconstructs the waveform while altering the fine structure, whereas additive noise acts differently, degrading phase-locking in the low-frequency channels. The removal chain also alters temporal fine structure through tempo and pitch changes and yields near-zero survival for all four baselines~\cite{oreilly2025shallow}.

Future work will target resistance to removal by carrying the key redundantly across bands and time scales, raising the degradation a remover must introduce. The same redundancy is expected to extend the usable range under low-bitrate neural codecs. A further direction is capacity: the present detector reports watermark presence, and extending the keyed projection to a multi-bit payload would allow the identifier itself to be recovered.

\section{Conclusion}

CochlearMark is a training-free audio watermark built on an auditory-nerve phase-locking representation. A keyed random projection serves as both the detector and the embedding target, without training an encoder or detector. CochlearMark achieves imperceptibility and signal-edit robustness comparable to the evaluated baselines, while its behavior under neural-codec re-encoding and content-altering transformations follows the preservation of temporal fine structure.


\clearpage
\small\bibliographystyle{IEEEbib}
\small\bibliography{strings,refs}

\end{document}